\documentstyle[sprocl,epsfig,epsf]{article}

\def\PRD{{\em Phys. Rev.} D}

\def\be{\begin{equation}}
\def\ee{\end{equation}}
\def\bea{\begin{eqnarray}}
\def\eea{\end{eqnarray}}

\begin{document}
\begin{flushright}
CERN-TH/96-164
\end{flushright}
\vspace{1.0cm}
\title{INTERFACES IN HOT GAUGE THEORY}

\author{S. BRONOFF }

\address{Centre Physique Th\'eorique au C.N.R.S., B.P. 907,\\ Campus de Luminy,
 F13288 Marseille CEDEX, France}

\author{C.P. KORTHALS ALTES}

\address{Theory Division, CERN, CH-1211 Geneva 23 Switzerland,\\ and 
Centre Physique Th\'eorique au C.N.R.S., B.P.
907, \\ 
Campus de Luminy, F13288 Marseille CEDEX, France}

\vspace{1.0cm}
\maketitle
\abstracts{ Consider Hamiltonian Z(2) gauge theory in a heat bath with temperature T. The
stringtension at low T and the free energy of domainwalls at high T can be computed  from one and the same
observable.  We show by explicit calculation that domain walls in  hot Z(2) gauge theory have good
thermodynamical behaviour. This is due to roughening of the wall, which expresses the restoration of
translational symmetry.}
\vspace{1.0cm}
\begin{center}
Contribution to the Proceedings of the Second Workshop on\\
``Continuous Advances in QCD"\\
28--31 March 1996, Minnesota, U.S.A.
\end{center} 

\vspace{2.0cm}

\begin{flushleft}
CERN-TH/96-164\\
June 1996
\end{flushleft}
\vfill\eject

\section{Introduction} Traditionally, spontaneously broken symmetries are restored at high enough temperature.
High enough means the scale of the energy gap in the theory. Nature provides us with such systems abundantly,
such as superconductivity, ferromagnetism and the electroweak sector of the standard model of particle physics.

 Much less known is the fact that not only nature, but also the laboratory of Lagrangian field theory, provides
us with systems where the opposite happens. Weinberg~\cite{wei} quotes the example of Seignette salt, a
ferroelectric compound; below the critical temperature it orthorombhic and above only isoclinic.
 Also, during the last two decades examples in field theory have been given mainly in connection with particle
physics and cosmology.

Independently of these developments, the study of SU(N) gauge theories has led to the notion of spontaneously
broken centergroup (Z(N)) symmetry~\cite{sveyaf} taking place above the deconfinement transition. This notion has
been useful in understanding certain universal properties of the transition, like critical exponents~\cite{gav}.
 
Other consequences of the spontaneous breaking are domain walls between coexisting phases. These questions have
been addressed at very high temperatures where semi-classical techniques are possible~\cite{bhat}. It was soon
discovered that these walls showed thermodynamic anomalies~\cite{bel}; and other criticism sprang up, going as
far as to deny the concept of broken Z(N) symmetry.
 
In this paper we want to restore the balance and point out that, in simple gauge systems, a fully controlled
calculation gives perfectly sensible answers to thermodynamical behaviour of the walls. The model in question is
Z(2) gauge theory, formulated with a lattice Hamiltonian in two space dimensions. The reason one has control is
that the model can be reformulated as a two-dimensional Ising model. Its three-dimensional version is not
analytically solvable, but  similar properties hold.
 
Although the reader may think Z(2) theory a little outlandish, we believe the results will carry over to an
SU(2) theory, because the main actors are electric fluxes and are present in  both theories.
 
In the next section we will formulate the problem in a general way. The subsequent section deals with the Z(2)
case. Most of Sections~2 and 3 has been known for 15 years. Only from Sec.~4 on are there developments which
to the best of our knowledge have not appeared in the litterature: a more precise formulation  of boundary
conditions, and the roughening of the surface.

\section{Formulation of the Problem } Let us start with some simple observations on a domain wall in between two
ordered phases.
 
If we have a system that breaks the symmetry at low temperature, the wall reflects the properties of the high
temperature phase. This implies that  entropy and internal energy of the wall are higher than those in the
surroundings. Needless to say, the free energy of the wall is {\it{higher}} because otherwise the system would
be unstable against wall formation.
 
Now the case under study: the symmetry breaks at high temperature. Then the wall has the properties of the low T
phase- hence a lower entropy and internal energy! Of course stability requires that the resulting free energy be
higher, just as in the previous case.
 
Local free energy and entropy are not unambiguously defined inside the wall.   Only the internal energy is
locally defined as the expectation value of the local Hamiltonian.

The free energy $F$ of a gauge theory is related to the Gibbs sum $Z$ over physical  states:
\be
\exp{-{F\over T}}\equiv Z=\sum_{phys}\langle phys\vert\exp{-H\over T}\vert phys\rangle .
\ee
 
A physical state in a gauge theory can be written as the average over all gauge transformations $\Omega(\vec x)$:
$$\vert phys\rangle =\int D\Omega\vert \vec A^{\Omega}\rangle .\eqno (2)$$
 
We suppose that boundary conditions on $\Omega$ have been imposed, such that any product of two of them again
obeys the boundary conditions.
 
With this definition a physical state is gauge-invariant. A gauge transformation in Hilbert space is generated
by the Gauss operator $G(\vec x)\equiv \vec D\cdot\vec E(\vec x)$ in the following way:
$$\vert \vec A^{\Omega}\rangle=\exp{i\omega\cdot G}\vert\vec A\rangle\eqno (3)$$ 
The dot means
summation over all indices, such as space and colour and $\Omega\equiv \exp{i\omega}$. The Gauss operator commutes with the Hamiltonian, and the
gauge invariance of a physical state leads to:
$$\langle\vec A^{\Omega}\vert\exp{-H\over T}\vert phys\rangle = \langle\vec A\vert\exp{-H\over T}\vert
phys\rangle .\eqno (4)$$ 
In other words, the presence of the physical state renders the gauge transformation on
other $\vert \vec A\rangle$ states redundant.  This is an important property and
corresponds, in the Euclidean path integral version of the free energy, to the freedom in dropping integrations
over  
$A_0$ variables in the Euclidean time direction without changing the free  energy.
 
 Here we want to avoid Euclidean path integrals altogether. Let it suffice to mention that in the path integral
one can define the Polyakov loop as the ordered product of all the $A_0$ potentials along the line in the
Euclidean time direction from 0 to $1\over T$. This formulation is useful numerically, but is obscuring the
issues at hand, i.e. the formation of surfaces in space.  Due to the above invariance  the Wilson line can be
reduced to the sole gauge transform $\Omega$. Nowhere  will we introduce Euclidean time.

 Instead, we will combine Eqs.~(1), (2) and (4) and do the integrations over $\vec A$ to get an effective action
in the {\it {same}} dimensionality as the original Hamiltonian:
 
$$\int D\vec A\langle\vec A\vert\exp{-H\over T}\vert\vec A^{\Omega}\rangle
\equiv\exp{{-S_{eff}}}\eqno (5)$$ so that 
$$\exp{-F\over T}=\int D\Omega\exp{-S_{eff}(\Omega)}.\eqno (6)$$ The gauge transformation lives in the same
space as that of the quantum states and the Hamiltonian.   The effective action describes the interactions
between the gauge group variables. There is no Euclidean time in Eq.~(6).
 
What purpose does the effective action serve?

It defines  averages such as $\langle\Omega(\vec 0\rangle$ (the average of the Wilson line) and more generally
correlations. These correlations are, in the language of the quantum statistical Gibbs sum (8), simply the
correlations of heavy charges~\cite{macsve}. These correlations teach us about the forces present in the system
for various temperatures. The import of the effective action is now clear, and in Sec.~3 these correlations
are worked out for the simple Z(2) model.

 \section{The  Z(2) Gauge Model}
\subsection{The Model} We study a  system  with gauge group Z(2) on a spatial lattice with continous   time at
equilibrium with a thermostat at temperature T. We will partly follow  Kiskis~\cite{kis} The Hamiltonian reads
$$H=\sum_lK(1-\sigma_x(l))   ,\eqno (7)$$
$K$  is the coupling and
 $\sigma_x(l)$ is the first Pauli matrix defined on the link $l$. It acts as such on the two-component wave
function in the link $l$ and acts as unity on wave  functions on the other links. The system is represented by
the product of those two-component $\psi$'s:
$$\Psi=\Pi_l\psi(l)\;.$$

 The Hamiltonian contains only  a kinetic term analogous to an electric term in an $SU(N)$
gauge theory. We omit the magnetic plaquette term for simplicity. It is not essential to the argument of this
paper. 
 
The ensuing theory is not trivial! The partition function Z in Eq.~(1) has physical states satisfying  Gauss's
law
$$G(\vec x)\vert phys\rangle\equiv\Pi_{\{l\gets l_{\vec x}\}}\sigma_x(l)\vert phys\rangle=\vert phys\rangle\eqno
(8)$$ 
We can diagonalize all the $\sigma_x$ and obtain on any link an eigenvalue $ 2K$ or $0$. We then have 
$$Z(T/K)=\sum^{\prime}_{\{\phi(l)\}}\Pi_l\{\exp{-{2K\over T}\phi(l)}\},\eqno (9)$$ 
where the prime denotes the
Gauss's constraint and $\phi(l)$ takes the values 0 or 1. If the latter we say there is ``flux'' on link l.
 
It is easy to see that a physical state can be represented by a configuration of fluxes such that an even number
enters into every vertex of the lattice. One can define the charge
 $$Q(\vec x)=\sum_{\{l\gets l_{\vec x}\}}\phi(l)\equiv \Phi(\vec x),\b mod 2 \eqno (10)$$ so Gauss's law in Eq.~(8) does not allow charge.

\subsection{Order Parameter}

We will now establish the relation between correlations of Wilson lines and the correlations between heavy test
charges. 
  First we need states with a test charge at $\vec y$. To this end we introduce  the two projectors :
$$P_e(\vec x)\equiv{ 1\over 2}\sum_{\omega(\vec x)=0, 1}G(\vec x)^{\omega(\vec x)}\eqno (11a)$$ and 
$$P_o(\vec x)\equiv{1\over 2}\sum_{\omega(\vec x)=0,1}(\exp{i\pi\omega(\vec x)})G(\vec x)^{\omega(\vec x)}\eqno
(11b)$$ We introduced here the notation $$\Omega(\vec x)\equiv\exp{i\pi\omega(\vec x)}=\pm 1, \eqno (12)$$ using
$G(\vec x)$ as defined in Eq.~(8).
 
In the basis where the Hamiltonian and hence the Gauss operator are diagonal, the projectors read:
$$P_e(\vec x)={1\over 2}\sum_{\Omega(\vec x)=\pm 1}\Omega(\vec x)^{\Phi(\vec x)}\eqno (13a)$$ and
 $$P_o(\vec x)={1\over 2}\sum_{\Omega(\vec x)=\pm 1}\Omega(\vec x)^{\Phi(\vec x)+1}.\eqno (13b)$$ The first projector
renders the number of fluxes ending in $\vec x$ even , the second one admits only an odd number, so creates a
charge mod 2. The projectors obviously commute with the Hamiltonian.
 
Consider a quantum state of the gauge field, which is physical everywhere except at $\vec y$, where an odd
number of fluxes ends (so where the charge sits). It can  be written with the help of the projectors as:
$$P_o(\vec y)\Pi_{\vec x\ne\vec y}P_e(\vec x)\vert\vec A\rangle\equiv \vert Q(\vec y),\vec A\rangle .\eqno (14)$$
 
The free energy $F_Q$ of the system with a charge is defined by:
$$\exp{-{F_Q\over T}}=\sum_{\{\vec A\}}\langle \vec A, Q(\vec y)\vert \exp{-{H\over T}} \vert \vec A, Q(\vec
y)\rangle .\eqno (15)$$ The projectors on the bra can be commuted through the Boltzmann factor and  give, with
Eqs.~(11):
 
$$\exp{-{F_Q\over T}}=\sum_{\{\Omega\}}\langle \vec A\vert\exp{-{H\over T}}\Omega(\vec y)\vert \vec
A^{\Omega}\rangle$$ Together with the definition of $S_{eff}$ in Eq.~(5) we find:
$$\exp{-{F_Q\over T}}=\sum_{\{\Omega\}}\Omega(\vec y)\exp{-S_{eff}}\eqno (16)$$
 
The reader will have noticed that an important point has been glossed over. It is the fact that a state with
only one charge in some point $\vec y$ is hard to realise, with Gauss' law satified everywhere else. In terms of
the fluxes: imagine one flux leaving $\vec y$ ( never to return there anymore) it has necessarily an end point
(or rather on odd number of them). These end points must be on the boundary, if we want  Gauss`s law everywhere
inside our volume. Thus we need at least one charge on the boundary, on which the flux can land. More generally
we need an odd number of charges on the boundary.
 So Gauss's law  forces us in  computing{\it{ correlations}} of an even number of Wilson lines, and the free energy $F_Q$ is obtained exponentially fast with distance.  

\section{Effective Action and the Ising Model}
 We start from Eq.~(9). We can undo the Gauss constraint on the sum by introducing
the projectors of Eqs.~(13) and find for the partition function:

$$Z(T/K)=2^{-V}\sum_{\{\Omega\}}\Pi_l\exp{-2{\phi(l)\over T}}\Pi_{\vec x}(\Omega(\vec x))^{\Phi(\vec
x)}.\eqno (17)$$
$V$ and $L$ stand in this section for number of points and number of links on the lattice.

The summation over all gauge field configurations $\{\phi(l)\}$ is now easily done and leaves us with:
$$\exp{-S_{eff}}=\Pi_l\big(1+(\Omega\Omega^{\prime})_l\exp{-2{K\over T}}\big).\eqno (18)$$ In this equation only products
of nearest-neighbour gauge transforms appear because a given flux $\phi(l)$ in Eq.~(10) is present as power in both, in Eqn.~(17).
 
Clearly the Ising model has emerged up to some factors. We find with the relation $$\exp{-2{K\over T}}=\tanh
J\eqno (19)$$ that the r.h.s. of Eq.~(17) can be rewritten as
$$
\exp{JL}{1\over {\cosh J}^L}\Pi_l\exp{-J(1-(\Omega\Omega^{\prime})_l)}.\eqno (20)$$  The relation between the
parameters can be extended to a local one: Eq.~(19) is true linkwise. Note that the dual coupling $J^*$ in the
Ising model equals
${K\over T}$.

This is the main result of this section. As expected from the previous section  we have  at high T (large J)
spontaneous breakdown of the Z(2) symmetry.

We note in Eq.~(17) that the Boltzmann weight of the gauge fields is accompanied by a factor depending on
$\Omega(\vec x)$ which is 
 negative when there is a charge in $\vec x$ (only after integrating over the gauge transform do we recover the
projectors, which give of course
 non negative factors). The   thermodynamical behaviour of the gauge fields in terms of $S_{eff}$ can therefore 
be quite unexpected. On the other hand the density matrix in terms of the variables $\Omega$, Eq.~(20) is 
positive.

Note that the behaviour is dual; it is obvious that at high $T$ many quantum states contribute to the partition
sum, whereas at large $J$ few classical spin configurations contribute, So the quantum entropy of the gauge
system is high whilst  the classical entropy of the Ising model is low.  
 We know that the Ising model  allows coexistence of different phases, separated by domain walls. One can
therefore ask how this translates into the gauge model at high temperature.

\section{Coexisting Phases}
There is a well-known device to create a domain wall in spin models: consider the boundary spins.
Fix the spins on one side to be up, on the other side to be down (see Fig.~1).
 
\begin{figure}
\hglue3cm
\epsfig{figure=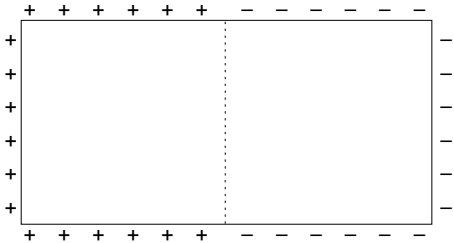,width=6cm}
\caption{The system with boundary spins frozen, such that a domainwall develops in the middle. Multiplying all
spins on one side of the wall with $-1$ will generate a dislocation along the centerline, and render the
boundary conditions homogeneous.
\label{fig:radish}}
\end{figure}

In the Ising model we have to introduce an external field $h_b$ at the boundary points b, which freezes the
spins in question.  Its counterpart in the gauge model  is a chemical potential $\mu_b$
that couples to the charge $Q(b)$ of Eq.~(10).  The relation between the  two is like that between the link
couplings in Eq.~(19):
$$\exp{-{\mu_b\over T}}=\tanh h_b\eqno (21).$$

Freezing spins with a positive (negative) field means $\mu_b=0^+ (0^++i\pi)$.
 It is clear that at large $J$ (high $T$ in the gauge model) the spins on both sides of this wall will have
opposite sign.  We have created a domain wall. To detect the wall one can put a charge on one side, and measure
the correlation with another charge. When going through the wall the correlation will go through zero and change
sign when emerging on the other side.

Of course, the configuration of chemical potentials  implies that any flux line passing  through the dislocation
an {\it{odd}} number of times will contribute with a minus sign to the partition function $Z_{+-}$! This is so
because such a flux will necessarily end with an odd number of end points on the boundary on both sides of the
wall, picking up an odd number of minus signs. This is perhaps easier to see by transforming all the spin
variables on one side of the wall with a minus sign. Then the border spins are all frozen the same way, and
there is no minus sign from the border. But now it comes from the antiferromagnetic dislocation ($J\rightarrow
-J$) we created by the transformation on the spins and by using the relation (19)  between $J$ and $T$!  Below
the critical $T_c$ there is no wall. Correspondingly, the correlation drops exponentially fast with distance and
is controlled by the string tension $\rho(J)$.
 
The ratio of the partition functions with and without mixed boundary conditions  signals in what phase we are:
\begin{eqnarray*} {Z_{+-}\over{Z_{++}}}&=&1 - \exp{-\rho(J)L}\mbox{\ \ \ if \ \ \ } J < J_c\\
               &=&\exp{-L_{tr}J\alpha(J)}\mbox{\ \ \ if \ \ \ } J > J_c 
\end{eqnarray*} The string tension has good thermodynamical properties~\cite{shig}. We now analyse those of the
domain wall.
\section{Domain Wall and Roughening} 
In the Ising model for $d=2$ the total free energy $\alpha(J)$ of the wall
is known  and equals $\alpha(J)=2(1-{J^*\over J})$ for $J\ge J^*$. When the two are equal, we have the Curie
point, where the surface stops to exist.
 
The location of the wall is subject to large fluctuations, of the order of $\sqrt L_{tr}$ for $d=2$. This is
known as roughening and its  consequence is that the scale over which the profile is varying becomes $\sqrt
L_{tr}$ in lattice units. The same is true for the energy density profile of the wall  in the Ising model. The
result is known analytically~\cite{abra} and reads:
$$
\epsilon_I(l)={a\over{\sqrt L_{tr}}}\exp{\{-\sinh(2(J-J^*)){z^2\over {L_{tr}}}\}}.
\eqno (22)$$ We will write ${z^2\over{L_{tr}}}=\zeta^2$. The link $l$ is supposed to
 run from  $z$ to $z+1$. Important is to note that the distance over which the interface is varying is of the
order of $\sqrt L_{tr}$. As the integrated profile gives us the same total energy, the amplitude of the energy
profile {\it{decreases}} like $\sqrt L_{tr}$. 
 
This formula reflects the random walk our one dimensional surface is making. It is universal up to the  typical
Ising factor in front of the exponent. The constant $a$ is such that upon integration we get the total energy
$2(1-{\partial J^*\over{\partial J}})$. To translate this result into the average $\epsilon(\zeta)$ of the 
energy density $1-\sigma_x(l)$ in the gauge model is straightforward. The result is:
$$ {\epsilon(\zeta)=\epsilon_0(T) - (-{\partial J\over {\partial J^*}})\epsilon_I(T)}.
\eqno (23)$$ The first term is the link energy density of the ground state of the gauge model, and is simply
 the Ising ground state energy per link, which is analytically known~\cite{abra}.

The Jacobian- reflecting the change from the variable J to the variable $KT^{-1}=J^*$- is always negative so the
energy density inside the wall is smaller than outside. This is one of the hallmarks of a broken symmetry at
high T, as we argued in the introduction. In Figs.~2 and 3 we show the dependence on $L_{tr}$.

\begin{figure}
\epsfig{figure=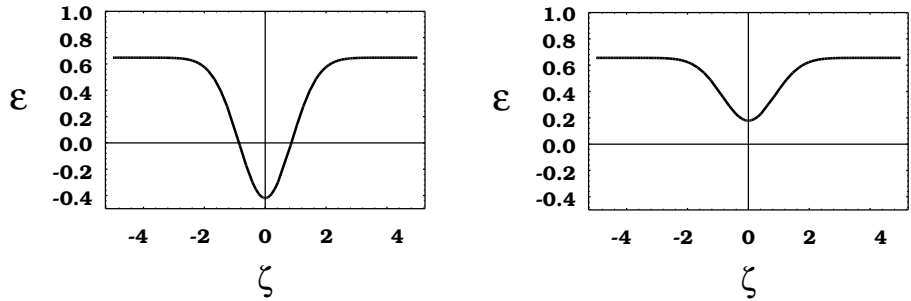,width=12cm}
\begin{minipage}[t]{.49\textwidth}
\caption[]{$\epsilon(\zeta)={\langle 1-\sigma_x(l)\rangle}(\zeta)$ is the mean value of the local energy operator
at physical temperature $T$ with $K/T=0.3$, Eqn.~(23). $\zeta$ is the rescaled variable ${z/ \sqrt{L}})$. The length of the wall is $L_{tr}=5$ in lattice units.
\label{fig:radish}}
\end{minipage}\hfill
\begin{minipage}[t]{.49\textwidth}
\caption{As in Fig.~2 but now for $L_{tr}=25$.
\label{fig:radish}}
\end{minipage}
\end{figure}
 

\section{Conclusions}
 
Domain walls and string tension in gauge theory can be described by the asymptotic properties of the same
partition function, for different temperature regimes. When the system is large enough  both string tension and wall have reasonable
thermodynamical properties, the  latter thanks to roughening. Its origin is the restoration of the translation
invariance broken by the interface. In $d=3$ the $\sqrt L_{tr}$ behaviour becomes logarithmic, and is only
present for a range of temperatures not too far above $T_c$.  

If one considers bubbles of radius R, the width of the wall will grow  as $\sqrt R$ or as $\log R$. The
total energy of the wall profile  depends only on temperature.

\section*{Acknowledgments}  We are indebted to Per Elmfors, Keijo Kajantie, Misha Shaposhnikov, Misha Stephanof,
and Mike Teper for many useful discussions and criticism. One of us (CPKA) thanks the organizers of this very
inspiring meeting and the CERN Theory Division for hospitality, S.B. acknowledges an Allocation de Recherche
MESR.

\end{document}